\documentclass[11pt]{article}
\usepackage[utf8]{inputenc}
\usepackage{amsmath}
\usepackage{amssymb}
\usepackage{amsthm}
\usepackage{graphicx} 
\usepackage{url}      
\usepackage{fullpage} 
\usepackage{hyperref}
\hypersetup{
    colorlinks=true,
    linkcolor=blue,
    citecolor=blue,
    filecolor=magenta,
    urlcolor=cyan
}
\usepackage[numbers]{natbib}   
\usepackage{longtable} 

\title{Computational Verification of the Buratti--Horak--Rosa Conjecture \\ for Small Integers and Inductive Approaches}
\author{
    Ranjan N Naik, Department of Mathematics, Lincoln University, PA, USA%
}
\date{July 30, 2025} 

\begin{document}

\maketitle

\begin{abstract}
This paper presents a comprehensive computational approach to verify and inductively construct Hamiltonian paths for the Buratti--Horak--Rosa (BHR) Conjecture. The conjecture posits that for any multiset $L$ of $p-1$ positive integers not exceeding $\lfloor p/2 \rfloor$, there exists a Hamiltonian path in the complete graph $K_p$ with vertex-set $\{0, 1, \dots, p-1\}$ whose edge lengths (under the cyclic metric) match $L$, if and only if for every divisor $d$ of $p$, the number of multiples of $d$ appearing in $L$ is at most $p - d$.

Building upon prior computational work by Mariusz Meszka, which verified the conjecture for all primes up to $p=23$, our Python program extends this verification significantly. We approach the problem by systematically generating frequency partitions (FPs) of edge lengths and employing a recursive backtracking algorithm. We report successful computational verification for all frequency partitions for integers $p < 32$, specifically presenting results for $p=31$ and a composite $p=26$. For the composite number $p=30$, the Python code took approximately 11 hours to verify on a Lenovo laptop. For $p=16$, $167,898$ valid multisets were processed, taking around 20 hours on Google Colab Pro+.

Furthermore, we introduce and implement two constructive, inductive strategies for building Hamiltonian paths: (1) increasing the multiplicity of an existing edge length, and (2) adding a new edge length. These methods, supported by a reuse-insertion heuristic and backtracking search, demonstrate successful constructions for evolving FPs up to $p=40$. Through these empirical tests and performance metrics, we provide strong computational evidence for the validity of the BHR conjecture within the scope tested, and outline the scalability of our approach for higher integer values.
\end{abstract}



\section{Introduction}

The study of Hamiltonian paths and cycles in graphs is a fundamental area within combinatorics and graph theory. A particularly challenging class of problems involves constructing Hamiltonian paths with prescribed properties, such as specific edge lengths or differences between consecutive vertices in graphs defined over cyclic groups. The Buratti--Horak--Rosa (BHR) Conjecture, which posits the existence of Hamiltonian paths with specific length sequences in complete graphs whose vertices are elements of $\{0, 1, \dots, p-1\}$ (for integers $p$), is one such notable problem that has attracted considerable attention \cite{horak2009problem, pasotti2014buratti, pasotti2014new}.

Let $K_p$ be the complete graph on $p$ vertices labeled $0, 1, \dots, p-1$. The \emph{cyclic length} of an edge between vertices $x$ and $y$ is defined as:
\[
\ell(x, y) = \min(|x - y|, p - |x - y|).
\]
Given a multiset $L$ of $p-1$ positive integers, the BHR conjecture states that such a multiset $L$ is the list of edge lengths of a suitable Hamiltonian path in $K_p$ if and only if a necessary and sufficient condition, known as the \textbf{divisor condition}, holds:
\begin{quote}
For every divisor $d$ of $p$, the number of elements in $L$ divisible by $d$ must be at most $p - d$.
\end{quote}

Previous efforts to address this conjecture have included theoretical constructions and computational searches. Mariusz Meszka, as reported in \cite{horak2009problem}, made significant progress by computationally verifying the conjecture for all primes up to $p=23$. This paper aims to further advance these computational results by presenting a Python program designed to systematically search for such Hamiltonian paths. We provide new computational evidence that extends the verification of the conjecture to all integers $p < 32$, specifically confirming its validity for $p=31$. Our approach, unlike some prior work, considers one multiset per frequency partition. For composite $p=30$, the program verified $78,378,960,360$ valid multisets, with $4960$ multisets passing coprime conditions and the BHR conjecture. For $p < 31$, the Co-prime BHR conjecture is verified.

Additionally, this paper introduces and validates two constructive, inductive strategies for generating Hamiltonian paths corresponding to specific edge frequency profiles. These methods, implemented algorithmically, provide a practical framework for exploring the conjecture and generating realizations for admissible multisets, extending the empirical validation up to $p=40$. The program's design emphasizes efficiency and scalability, allowing for its utilization on more powerful computing resources to explore even higher integer values, thereby providing further evidence for the validity of the conjecture.

\section{Methodology}

The core of our computational approach is a Python program designed to construct Hamiltonian paths based on specified "frequency partitions" (FPs) of edge lengths. For a given integer $p$, the vertices of the complete graph are represented by elements of $\{0, 1, \ldots, p-1\}$. The "edge lengths" (or "hops") are defined as the absolute differences between connected vertices modulo $p$. These lengths are typically taken from the set $\{1, 2, \ldots, \lfloor p/2 \rfloor\}$.

A frequency partition (FP) represents a multiset of these edge lengths such that their sum equals $p-1$. Each FP corresponds to a potential set of "hops" that could form a Hamiltonian path of length $p-1$ (connecting $p$ vertices). The program proceeds as follows:

\begin{enumerate}
    \item \textbf{Frequency Partition Generation:} The \texttt{generate\_fps(n)} function systematically generates all possible frequency partitions for a given sum \texttt{n} (which is $p-1$ in our case) in lexicographic descending order. It ensures that the generated partitions adhere to relevant constraints from the divisors of $p$, such as the number of distinct hop values not exceeding $\lfloor p/2 \rfloor$.
    \item \textbf{Representative Multiset Conversion:} For each generated FP, the \texttt{get\_representative\_multiset(fp\_tuple, p\_val)} function converts it into an actual multiset of hop values. For instance, an FP like $(2, 1, 1)$ for $p=8$ might correspond to the multiset $\{1, 1, 2, 3\}$.
    \item \textbf{Recursive Path Search with Inductive Strategies:} The \texttt{\_recursive\_sequence\_builder} function implements a depth-first search (DFS) with backtracking to find a Hamiltonian path.
    \begin{itemize}
        \item It starts from a fixed initial vertex (e.g., node 0, \texttt{path=[0]}).
        \item At each step, it attempts to extend the current path by choosing an available hop from the multiset of edge lengths.
        \item A hop $h$ connects the current node $u$ to either $(u + h) \pmod{p}$ or $(u - h) \pmod{p}$. The search prunes branches where the next node has already been visited, or where no valid hop can extend the path.
        \item The \texttt{remaining\_counts} dictionary efficiently tracks the availability of each hop.
        \item A \texttt{hint} mechanism, which reuses successful permutations from previous frequency partitions, is incorporated to potentially speed up the search for subsequent partitions. This hints at the inductive nature of the search.
    \end{itemize}
    \item \textbf{Inductive Construction Scenarios:} Beyond brute-force searching all FPs, the system also employs inductive strategies to build Hamiltonian paths for increasing values of $p$ or modified FPs. These strategies leverage previously found paths to construct new ones:
    \begin{itemize}
        \item \textbf{Scenario I: Increasing the Multiplicity of an Existing Part (K fixed):} We fix the number of distinct edge lengths $K$ and increase the multiplicity of one of the existing edge lengths by 1. This results in a new multiset $L_2$ of size $p$ that differs from $L_1$ by a single count increment. The primary method for this is \textbf{Reuse-Insertion}: attempting to insert the new vertex into the previously found Hamiltonian path and validating against the updated FP.
        \item \textbf{Scenario II: Increasing the Number of Parts (K to K+1):} We begin with a multiset $L_1$ of size $p-1$ with $K$ distinct edge lengths and construct a new multiset $L_2$ of size $p$ by adding one new edge length not present in $L_1$. This corresponds to increasing the number of parts in the fingerprint representation of the multiset. For this scenario, \textbf{Heuristic Scoring} is also employed to try top-scoring insertions that minimize FP deviation, followed by a \textbf{Backtracking Search} as a fallback.
    \end{itemize}
    \item \textbf{Verification Loop and Logging:} The \texttt{verify\_bhr\_for\_p()} (a generalized version of the original \texttt{verify\_bhr\_for\_p29()}) iterates through all generated FPs or inductively evolved FPs. For each FP, it calls the \texttt{\_find\_sequence} function, which wraps the recursive search and manages global state for tracking success, the found path, and the number of backtracks. Each run is logged with timestamps, methods used, FP and HP details, backtrack counts, and computation time. This detailed logging provides a verifiable record of the search process.
\end{enumerate}

The program's design ensures a comprehensive exploration of all valid frequency partitions for a given integer $p$, systematically checking for the existence of corresponding Hamiltonian paths, and leveraging inductive methods for efficiency.

\section{Results}

Our computational experiments successfully verified the conjecture for all frequency partitions for integers $p < 32$. Specifically, we present compelling results for $p=31$ and, for illustrative purposes, a composite number $p=26$. We also include results from the inductive construction approach up to $p=40$.

For \textbf{$p=31$}:
The program identified a total of \textbf{5096 distinct frequency partitions}. For every single one of these 5096 partitions, the program successfully found a corresponding Hamiltonian path. The search for each individual frequency partition was remarkably fast, often completing within 0.00 to 0.03 seconds on a standard computer. The number of backtracks for finding solutions ranged from 0 for simple cases (e.g., $\text{FP}=(30)$) to tens of thousands for more complex partitions (e.g., $\text{FP}=(2, 2, 2, 2, 2, 2, 2, 2, 2, 2, 2, 2, 2, 2, 2)$ required 19206 backtracks). The total time taken for the comprehensive verification of all 5096 FPs for $p=31$ was highly efficient, indicating the effectiveness of the algorithm.

A sample of the successful verification logs for $p=31$ is shown below:
\begin{verbatim}
=== BHR Conjecture Verification for p = 31 ===
[07:09:57] Starting search for p=31, total 5096 FPs
[07:09:57] Path found for FP=(30,) in 0.00s with 0 backtracks
Permutation: [1, 1, 1, ..., 1]
Path: [0, 1, 2, ..., 30]

[07:09:57] Path found for FP=(29, 1) in 0.00s with 5 backtracks
Permutation: [1, 1, ..., 1, 2, 1]
Path: [0, 1, ..., 28, 30, 29]
...
[12:06:44] Path found for FP=(2, 2, 2, 2, 2, 2, 2, 2, 2, 2, 2, 2, 2, 2, 2) in 0.03s with 19206 backtracks
Permutation: [1, 1, 2, 2, 3, ..., 13]
Path: [0, 1, 2, ..., 21]

[12:06:44] Total FPs fixed: 5096 / 5096
\end{verbatim}

For \textbf{$p=26$} (a composite number), the program identified \textbf{1763 frequency partitions}, all of which were successfully verified:
\begin{verbatim}
For p=26
[11:15:27] Path found for FP=(3, 2, 2, 2, 2, 2, 2, 2, 2, 2, 2, 1, 1) in 0.08s with 38145 backtracks
Permutation: [1, 1, 1, 2, 2, ..., 9]
Path: [0, 1, 2, ..., 6]
...
[11:15:27] Path found for FP=(2, 2, 2, 2, 2, 2, 2, 2, 2, 2, 2, 2, 1) in 0.01s with 1094 backtracks
Permutation: [1, 1, 2, 2, 3, ..., 12]
Path: [0, 1, 2, ..., 15]

[11:15:27] Total FPs fixed: 1763 / 1763
\end{verbatim}
These results support the previously reported computational verification up to $p=23$ by Mariusz Meszka. The consistency of successful path finding across all tested frequency partitions for these numbers provides strong computational evidence for the conjecture's validity within this range.

\subsection*{Inductive Construction Results up to $p=40$}
In one run spanning 10 iterations, the system evolved frequency partitions to $p = 40$, maintaining successful Hamiltonian path constructions throughout. Selected iterations demonstrating the inductive approach are shown below. Each result indicates the use of backtracking, the specific $p$ value and evolved FP, the constructed HP and its frequency, the number of backtracks, and the computation time.

\subsection*{Iteration 9}
\begin{itemize}
    \item \textbf{Timestamp:} 20:19:41
    \item \textbf{Method used:} Backtrack
    \item \textbf{p (vertices):} 39
    \item \textbf{Evolved FP:} \texttt{Counter(\{1:6, 2:6, \ldots, 16:1\})}
    \item \textbf{HP:} \texttt{[0, 1, 2, \ldots, 9]}
    \item \textbf{HP freq:} \texttt{Counter(\{1:6, 2:6, \ldots, 16:1\})}
    \item \textbf{Backtracks:} 9002
    \item \textbf{Time:} 0.03 sec
    \item \textbf{Result:} SUCCESS
\end{itemize}

\subsection*{Iteration 10}
\begin{itemize}
    \item \textbf{Timestamp:} 20:19:41
    \item \textbf{Method used:} Backtrack
    \item \textbf{p (vertices):} 40
    \item \textbf{Evolved FP:} \texttt{Counter(\{1:6, 2:6, \ldots, 17:1\})}
    \item \textbf{HP:} \texttt{[0, 1, 2, \ldots, 20]}
    \item \textbf{HP freq:} \texttt{Counter(\{1:6, 2:6, \ldots, 17:1\})}
    \item \textbf{Backtracks:} 10827
    \item \textbf{Time:} 0.03 sec
    \item \textbf{Result:} SUCCESS
\end{itemize}

\subsection*{Summary Table of Inductive Runs}
\begin{longtable}{|c|c|c|}
\caption{Backtracks for Inductive Runs} \\
\hline
Iteration & $p$ & Backtracks \\
\hline
\endfirsthead
\multicolumn{3}{c}{\tablename\ \thetable{} -- continued from previous page} \\
\hline
Iteration & $p$ & Backtracks \\
\hline
\endhead
\hline
\endfoot
\hline
\endlastfoot
1 & 31 & 1765830 \\
2 & 32 & 82254492 \\
3 & 33 & 5981815 \\
4 & 34 & 90111 \\
5 & 35 & 565485 \\
6 & 36 & 162158 \\
7 & 37 & 1297988 \\
8 & 38 & 2212 \\
9 & 39 & 9002 \\
10 & 40 & 10827 \\
\end{longtable}

These results from both direct verification of all FPs and inductive construction scenarios demonstrate the robustness of the BHR conjecture for the tested integer ranges. The inductive approach, while still relying on backtracking for challenging cases, offers a promising path for exploring larger $p$ values by leveraging previous successful constructions. For $p=30$, the program verified $78,378,960,360$ valid multisets and found that $4960$ multisets passed coprime conditions and passed BHR conjecture. For $p=16$, $170,544$ multisets were generated, which reduced to $167,898$ after applying the necessary condition of the BHR Conjecture. The verification process for $p=16$ took approximately 20 hours on Google Colab Pro+, highlighting the high memory and CPU consumption for larger $p$. For $p > 100$ with some minimal multisets, Hamiltonian paths were constructed efficiently, while for $p=55$, it took 11 digits of backtracks to complete on a Lenovo ThinkPad.

\section{Conclusion and Future Work}

We have presented a comprehensive computational verification of the Buratti--Horak--Rosa Conjecture for integers $p < 32$, specifically demonstrating its validity for $p=31$ and $p=26$ by successfully constructing Hamiltonian paths for all associated frequency partitions. This work extends prior computational efforts by Mariusz Meszka and provides robust support for the conjecture. Furthermore, we demonstrated the efficacy of inductive construction strategies, extending successful verifications up to $p=40$.

The Python program developed for this research is efficient for the tested ranges and is designed to be scalable. The integration of reuse-insertion heuristics alongside backtracking significantly enhances its performance for evolving frequency partitions. Future work will involve leveraging more powerful computing resources from Colab and exploring potential algorithmic optimizations to extend the verification to even larger integer values. The program's modular structure allows for straightforward adaptation to investigate related combinatorial problems concerning Hamiltonian paths and graph decompositions. The long computation times for larger $p$ values (e.g., $p=16, 30, 55$) underscore the complexity of the problem and the need for further optimization or specialized hardware.

\section*{Acknowledgments and Code Availability}
The authors acknowledge the prior work by Mariusz Meszka that laid the groundwork for computational verification of the BHR conjecture. The Python program used for this verification is available upon request to the authors (contact: johndoe@lincoln.edu).


\end{document}